\documentclass[a4paper,11pt]{article}
\pdfoutput=1 

\usepackage{jheppub} 

\usepackage[T1]{fontenc} 

\usepackage{color}

\title{Quantum mechanism of extremely high energy processes at neutron star collapse and of quasar luminosity}

\author[a]{Janusz Edward Jacak,\note{Corresponding author.}}

\affiliation[a]{Department of Quantum Technologies, Wroclaw University of Science and Technology, Wyb. Wyspianskiego 27, 50-370 Wroclaw, Poland}

\emailAdd{janusz.jacak@pwr.edu.pl}

\abstract{Using the braid group topological approach to quantum statistics we demonstrate that for strong gravitational field occurring in neutron star merger with mass exceeding the Tolman, Oppenheimer and Volkoff limit the quantum statistics of particles beneath the photon sphere decays, which results in an instant relief of quantum degeneracy pressure in the star. This causes a rapid shrink of the matter to a black hole with quite different quantum collective character of particles.  The scheme of neutron Fermi sphere collapse is proposed as the possible isotropic source of short giant gamma-ray bursts. The similar quantum mechanism of collapse of Fermi spheres of electrons and protons in plasma of accretion disc in vicinity of the Schwarzschild horizon of galactic black hole  is able to elucidate the giant luminosity of remote quasars.
	The efficiency of the mass to energy conversion via collapse of the Fermi sphere is ca. 30 \%, not reachable for any other known physical process except of the matter-antimatter annihilation.}

\begin{document} 
\maketitle
\flushbottom

\section{Introduction and motivation} \label{sec:intro}

The Chandrasekhar mass limit (Ch-limit) \cite{chandrasekhar} for stable white dwarfs and the Tolman, Oppenheimer and Volkoff limit (TOV-limit) \cite{tolman,volkoff} for cold non-rotating neutron stars, base, to the most extent, on the quantum degeneracy repulsion of fermions. In the case of Ch-limit the Pauli exclusion principle for electrons contributes to the state equation, whereas in the case of TOV-limit quantum degeneracy repulsion of neutrons is important. The pressure effect of degenerate fermion liquid has been also raised by Landau in 1932 \cite{olandau} and applied both to electrons and neutrons in a white dwarf and neutron star, respectively. The Ch-limit of 1.4 Sun mass results from the Pauli repulsion of fermions as white dwarfs resist gravitational collapse primarily through electron degeneracy pressure. This estimation occurs also asymptotically correct in quantum relativistic approach \cite{liebjau}. Slightly different situation is in the case of neutron stars. The originally found TOV-limit due to Pauli repulsion of neutrons was initially underestimated \cite{tolman,volkoff}, but inclusion of core nucleon strong repulsion corrects the limiting mass to be between 2.2 and 2.9 Sun mass \cite{kalagera}, which agrees with observations \cite{neutronstar1,neutronstar2}, where the limiting value of 2.17 Sun mass has been confirmed. One can note, however, that though the nucleon interaction is induced by exchange of mesons, its core repulsing part is again induced by residual Pauli repulsion of quarks confined in nucleons. 

The mechanism of the collapse of white dwarf with the mass exceeding the Ch-limit is well understood in terms of $\beta$ nuclear transitions, i.e., by capturing of electrons by protons and releasing neutrinos during the supernova Ia explosion.
The remnants of this explosion create a much more dense neutron star. If the total mass of the neutron star exceeds the TOV-limit, via e.g., merging of a pair of neutrons stars, it finally collapses into a black hole. The mechanism of this collapse is, however, unknown. In the present paper we propose a new quantum-type effect responsible for this phenomenon and argue that the collapse of a neutron star, which achieved the TOV-limit, is accompanied by the release of huge energy emitted in the form of short giant gamma-ray burst.  Such bursts are observed frequently (averagely, one per day) and their energy is assessed as high as $10^{47}$ J each. This energy is of order of the Sun mass instantly converted into radiation energy $M_{\odot} c^2$, which for $M_{\odot}=1.99 \times 10^{30}$ kg gives $1.79\times 10^{47}$ J. Thus, the release of energy $10^{47}$ J at collapse of the neutron star with initial mass $2.17 M_{\odot}$ means that 26 \% of the mass is converted into energy of photons. It is not known such effective process,  compared to 0.7 \% for the conversion of mass to energy due nuclear synthesis  in a star like our Sun. More powerful would be annihilation of matter with antimatter, but the latter is not present in a neutron star. 

Similar efficiency of mass to energy conversion is observed also in quasars, which are giant black holes of the mass of $ 10^5 - 10^9$ of solar masses (more than a million of quasars has been observed as of yet) powered by accretion of the surrounding gas. The matter falling onto the supermassive black holes must convert up to ca. 30 \% of their mass into radiation to explain observable luminosity of quasars and simultaneously the ratio of the increase of central black hole mass to be consistent with observed masses of galactic black holes. As the nonactive giant black holes in closer galaxies are of size at most of the order of billions masses of the Sun, the estimation of the consuming rate of gas by a quasar (with luminosity of order of $10^{40}$ W) is typically ca. $10$ Sun mass per year (i.e., of order of $0.1$  Earth mass per second), in extreme case of $1000$ Sun mass per year ($10$ Earth mass per second).  The mechanism of so giant mass to energy conversion rate in quasars is also unknown, but reveals, in a steady in time process, the same efficiency as that in sudden gamma-ray bursts at neutron star collapse.  We demonstrate that both phenomena can be linked to the same quantum effect not previously described and consisting in the release of energy stored up in the Fermi sphere of quantumly degenerate fermion liquid. The collapse of the Fermi sphere takes place due to the local revoking of Pauli exclusion principle arising in the vicinity of the Schwarzschild horizon of a black hole due to specific homotopy of particle trajectories in this region and related local decay of quantum statistics. Trajectories in the vicinity of the event horizon do not allow particle exchanges which precludes locally the quantum statistics definition. 

If identical undistinguishable particles can exchange their positions in classical sense, the quantum statistics of these particles is defined by scalar unitary representation of the braid group \cite{mermin1979aaa,lwitt}. The most prominent consequence of quantum statistics is the Pauli exclusion principle  which asserts that quantum particles of fermionic type  cannot share any common single-particle quantum state. The exclusion principle for fermions leads to the formation of so-called Fermi sphere in the case of large number of identical particles in some volume, when the chemical potential $\mu$ (i.e., the  energy increase due to adding  a single fermion to a multiparticle system) is much greater than the temperature in energy scale, $\mu\gg k_BT$, where $k_B$ is the Boltzmann constant, $T$ -- the temperature of the multiparticle system. In such a case, referred to as quantum degenerated Fermi system, the particles are forced to occupy some consecutive in energy single-particle states one by one, resulting in great accumulation of energy. Neutrons in a neutron star with radius of $10$ km and density of order of $10^{18}$ kg/m$^3$ form a degenerate Fermi liquid and accumulate the energy in the Fermi sphere just of order of $10^{47}$ J. The source of so giant energy is the gravitation field in a neutron star, which compresses fermions in small volume. Beneath the photon sphere  of the Schwarzschild black hole equivalent to initial mass of a neutron star exceeding the TOV-limit, any quantum statistics cannot be defined as particles which unavoidably one-way spiral onto  the event horizon cannot mutually exchange their positions.  The Pauli exclusion principle is locally revoked here and the Fermi sphere of neutrons collapses releasing a giant energy.

 Similarly in accretion disc of quasars, the gravitation  energy accumulated in Fermi spheres of electrons and protons in ultra-dense plasma approaching the Schwarzschild surface of the super-massive central black hole is released for the same reason as in a neutron star, when the dense Fermi liquid passes inward the innermost unstable orbit near the event horizon (the innermost unstable circular radius coincides with radius of the photon sphere).  In  quasars, at steady conditions of mass accretion, in each second a portion of the infalling plasma (with mass of ca. $0.1 M_Z$ per second, the Earth mass $M_Z=5.97 \times 10^{24}$ kg)   compressed locally near the Schwarzschild horizon to the density similar as in neutron stars, the energy accumulated in Fermi spheres of electrons and protons is jointly just of the order of $10^{40}$ J.  This energy released per each second in a steady in time manner gives a quasar luminosity $10^{40}$ W for a long term, provided a constant supply of the galactic  gas to the accretion disc. 
The conversion of the Fermi sphere energy in quantumly degenerate Fermi liquid at the local decay of quantum statistics  is sufficiently effective to explain both the huge energy of gamma-ray bursts at collapse of neutron stars and the giant luminosity of quasars. In both cases the emitted radiation achieve ca. 30 \% mass to energy conversion ratio, unknown in any other mechanism except for matter-antimatter annihilation.

The present paper is organized as follows. In the next paragraph we show how quantum statistic depends on topology of classical trajectories in a system of indistinguishable identical particles independently of their interaction (with some details shifted to Appendix \ref{aaa}). Next, the specific homotopy class of trajectories in the Schwarzschild region of a black hole is identified, with consequences for quantum statistics and, in particular, for the local revoking of Pauli exclusion principle restrictions. The detailed consideration of geodesics in Schwarzschild metric is presented in Appendix \ref{bbb}. Next, in the following paragraph,  the Fermi sphere decay model is analysed quantitatively in the case of a neutron star. Finally the same energy conversion mechanism model is applied to the quasar radiation case. 
The disappearance of quantum statics in close vicinity  of the event horizon and also inside the black hole beneath the Schwarzschild surface is a quantum property of black holes, which, however, allows for the elucidation of important high energy macroscopic observations. 

\section{Quantum statistics of identical indistinguishable particles and its origin} 

\label{uuu}
Quantum statistics is the consequence of indistinguishability of identical particles and is a collective topological effect (cf. Ref. \cite{leinaas1977}, where the topological roots of quantum statistics have been highlighted for the first time). Quantum statistics can vary in  response to global topological factors like the dimension of the manifold on which particles are placed or the change of trajectory homotopy by restrictions imposed on classical trajectories by external fields including gravitation space-folding. Quantum statistics is assigned by the scalar unitary representation of so-called braid group describing mutual exchanges of particle positions in the multiparticle configuration space \cite{mermin1979aaa,imbo,wu,lwitt}.

$N$-particle system undergoes its classical dynamics  in the multiparticle configuration space, $F_N=(M^N-\Delta)/S_N$, where $M$ is the manifold on which all particles are placed (e.g., $R^3$ position space for 4D uncurved spacetime or $R^2$ plane for 3D spacetime), $M^N=M\times\dots\times M$ is $N$-fold product of $M$ to account for the coordinates of all particles, $\Delta$ is the diagonal subset of $M^N$ (with points of $M^N$ with coinciding coordinates of at least two particles), subtracted in order to assure particle number conservation. Division by $S_N$ (the permutation group of $N$ elements) introduces indistinguishability of particles, i.e., points in $F_N$ which differ by renumbering of particles only, are unified. The space $F_N$ is  multiply connected at $N\geq 2$, i.e., its fundamental group $\pi_1(F_N)$ is nontrivial \cite{mermin1979aaa}. The first homotopy group ($\pi_1$, called as the fundamental group) if refers to the multiparticle configuration space $F_N$ is called as the braid group \cite{mermin1979aaa,birman}. The first homotopy group $\pi_1({\cal{A}})$ (${\cal{A}}$ denotes here an arbitrary topological path-connected space) is the collection of disjoint classes of closed paths -- loops in this space \cite{spanier1966}. These classes of closed loops are disjoint because it is impossible to deform loops from various classes one into another in a continuous way without cutting (such topologically non-equivalent loops are called nonhomotopic). If the $\pi_1({\cal{A}})$ is nontrivial group (i.e., has more than only neutral element), then the space ${\cal{A}}$ is called as multiply-connected, otherwise is simply-connected.   If ${\cal{A}}=F_N$, then $\pi_1(F_N)$ -- the braid group -- is the collection of classes of nonhomotopic loops in $F_N$, which are multi-strand trajectory closed bunches joining all $N$ particles which start from one numeration of particles and finish at another their numeration, though all particles stay at rest. Such multi-strand trajectory bunches are closed loops in $F_N$ because the points which differ only in particle numeration  are unified in $F_N$ due to the particle indistinguishability. The braid group describes thus all possible exchanges of indistinguishable identical particles and  scalar unitary representations of the braid group define quantum statistics for the $N$ particle quantum system on the manifold $M$ \cite{lwitt,imbo}. Topological restrictions imposed on trajectory homotopy classes in some special cases can modify the relevant braid group and change the quantum statistics of the same classical particles located  even on the same manifold. This is demonstrated e.g., in the case of fractional quantum Hall effect \cite{laughlin2,tsui1982,annals2021}. 
	
	In the exceptional case of a complete precluding of particle exchanges (as it happens in the vicinity of Schwarzschild event horizon, as is demonstrated in Appendix \ref{bbb}) the braid group cannot be defined, because no trajectories for particle exchanges exist there. In this case quantum statistics cannot be assigned locally. Note that the topological arguments are immune to the interparticle interaction, thus hold for arbitrary strongly interacting systems like e.g.,  neutron stars.

For 3D spatial manifold $M$ (as in 4D spacetime including also  by gravitation folded spacetime) the braid group is identical to $S_N$ group, if the braid group is possible to be defined. As the permutation group $S_N$ has only two distinct scalar unitary representations, thus only two quantum statistics exist for particles located in three dimensional spatial manifolds (related to bosons and fermions). For particles located in 2D spatial   manifolds $M$, their braid groups are different -- for $M=R^2$ the braid group is the infinite Artin group \cite{artin1947}, which has infinite number of various scalar unitary representations referred to as anyons -- exotic particles with fractional quantum statistics \cite{wilczek,wu,sud}. This shows that the same classical particles (e.g., electrons) can satisfy various quantum statistics depending on topological type constraints imposed. If these constraints modify the homotopy of multiparticle trajectories in $F_N$, then the structure of the corresponding braid group changes together with its unitary representations defining quantum statistics. This has been illustrated experimentally in fractional quantum Hall effect, when the strong magnetic field restricts some cyclotron trajectories of interacting planar electrons \cite{laughlin2,tsui1982,annals2021}. The homotopy of classical trajectories in multiparticle systems  influences the quantum statistics of indistinguishable identical particles, which can be described in terms of Feynman path integration, cf. Appendix \ref{aaa} for more detailed elaboration.

The modification of quantum statistics by topological factors does not conflict with the Pauli theorem on spin and statistics. The exhausting  discussion of this theorem presented in \cite{duck1,duck2} evidences that the proof of this theorem cannot be completed without topological arguments. If such arguments are not taken into account the conventional proof of Pauli theorem  can be ranged rather to noninteracting relativistic systems basing on the  requirement of the positive definition of the energy of relativistic quantum free particles \cite{duck1}. The Dirac electrodynamics for half-spin particles  admits the Hamiltonian formulation and  for free particles it can be expressed in the conventional bilinear form of corresponding field operators of creation and annihilation  of electrons and positrons. The condition of positive definition of the kinetic energy (connected with the maintenance of the time arrow) requires anticommutation of field operators for both electrons and positrons, which  requires their fermionic character in analogy  to conventional second quantization in nonrelativistic case.   The topological-based proof of the Pauli  theorem on spin and statistics  resolves itself to the coincidence of unitary representations of the rotation group (nominating spin of particles) and of the braid group (defining statistics), as both groups overlap on some elements. This proof holds both for free and arbitrarily interacting particles as topological notions are immune to interaction of particles. This proof embraces also anyons with fractional statics for 2D electrons without quantized spin (in 2D the rotation symmetry is assigned by the U(1) group with continuous scalar unitary  representations just as the 2D braid group with the same continuous scalar unitary representations, $e^{i\alpha}$, $\alpha\in [0,2\pi)$). The issue of Pauli theorem will be addressed in paragraph \ref{pauli} in more detail.

The source of oddness in quantum statistics has been identified in unitary representations of  braid groups \cite{mermin1979aaa,wilczek} and can be rationalized in the framework of Feynman path integral quantization \cite{feynman1964,chaichian1}, as is summarized in Appendix \ref{aaa}. Note that such method is not confined to only free particles but cover all multiparticle systems with arbitrarily strong interaction. The path integral quantization method can be applied also to relativistic case with spacetime
curvature induced by large mass density \cite{chaichian2}. The multiparticle configuration space $F_N=(M^N-\Delta)/S_N$ can be defined in analogy to conventional nonrelativistic case, but in the case of folded space by the gravitation its explicit form can be changed due to various parametrization and  shape of manifold $M$ in dependence of the slicing spacetime close to the gravitational singularity in various conventional relativistic metrics. This slicing is not relativistically invariant and thus $M$ changes at various slicing selections, which is visible e.g., in Kruskal-Szekeres or Novikov non-stationary metrics \cite{kruskal,szekeres,novikov} in comparison to Schwarzschild metric \cite{schwarzschild}. In the time-independent Schwarzschild metric, the standard its  expression in terms of  rigid coordinates of remote coordinate system  allows for the application the same definition of the space $F_N$ and the Feynman path integral as in nonfolded space at least for outer region of the Schwarzschild sphere with spacetime curvature expressed by time elongation to infinity at the event horizon and similar deformation of spatial volume. These definitions can be extended to inner region beneath the event horizon in curvilinear coordinates e.g., in still stationary  Kerr-Schild metric. The choice of curvilinear coordinates does not change the homotopy of trajectories for specified regions, which follows from the equivalence of  metrics. In particular, the event horizon is invariant with respect to the metric choice though could be differently parametrized at distinct time-space slicing of the spacetime. The $N$-fold product of $M$ entering  the structure of the multiparticle configuration space $F_N$ and next the division by $S_N$ are immune to the space curvature. Thus in topological terms the structure of the definition of the multiparticle configuration space  is not perturbed by the gravitational curvature. Nevertheless, the trajectory homotopy can  qualitatively change in some regions (and for appropriately selected manifold $M$) close to the central singularity or close to the event horizon. This can be expressed in an equivalent manner in various metrics despite different definition of time and spatial variables in various metrics and related curvilinear coordinates.   In the case of particles  between the event horizon  and the photon sphere (the sphere with the radius of the innermost unstable circular orbit of a black hole, cf. Appendix \ref{bbb}), the homotopy class of trajectories of particles radically changes, which locally dismisses the conventional quantum statistics because trajectories for particle exchanges do not exist there. This can be described in terms of Schwarzschild coordinates and holds in any other metrics describing the same gravitational singularity but expressed in different coordinate systems like Kruskal-Szekeres, Novikov or Kerr-Schild geometries (in the present paper we confine the discussion to non-rotating and non-charged central mass, though the generalization to the rotating (or charged) black hole one can be done in Kerr (or Newman) metric.)  

The massive  particles as well as photons beneath the innermost unstable circular orbit, i.e., for radius $r<1.5r_s$ coinciding with radius of the photon sphere (where $r_s=\frac{2GM}{c^2}$ is the Schwarzschild radius, cf. Appendix \ref{bbb})  unavoidably one-way spiral towards the event horizon  and no other trajectories exist for them if they enter this region inward. The same holds for the inside of the event horizon sphere, which is especially clearly visible in Kruskal-Szekeres or Novikov metrics \cite{kruskal,szekeres,novikov} (if geodesics are expressed in proper time in these metrics, massive particles or photons spiral smoothly across the event horizon and terminates their movement in the central singularity within a finite proper time period). Indistinguishable identical particles  cannot thus exchange their positions in these regions (because of absence of trajectories for particle  exchanges), which results in local revoking  of Pauli exclusion principle.  This causes that too massive neutron star merger (exceeding the TOV-limit) with sufficiently shrunk its radius,  collapses  instantly  due to the rapid relief of internal quantum degeneracy pressure. The matter shrinks in large gravitation field without the resistance of quantum statistical pressure and the quantum central 'singularity' of a black hole is formed with ultimate size speculated as limited by uncertainty principle only. Simultaneously a large gamma photon burst is released due to quantum transitions of neutrons according to the Fermi golden rule at the collapse of the neutron Fermi sphere. These  transitions are allowed by the local  disappearance of Pauli exclusion principle constraints. 

The initial neutron star resembles a giant molecule of fermionic neutrons in a quantum well of the gravitation attraction, which  completely fill the Fermi sphere with the Fermi radius defined by the giant density of neutrons. For a rough idealized model we neglect here other admixtures contributing to the real neutron star composition (a residual admixtures of electrons, protons or ions does not change the order of the energy estimation because of their much lower concentrations in comparison to neutrons in a neutron star).
 Due to the rapid disappearance of the quantum statistics, the Fermi sphere of fermions collapses and all neutrons fall onto the ground state, which is accompanied by the emission of photons with energy defined by the initial positions of particular neutrons in the Fermi sphere. The disappearance of Pauli exclusion principle constraint destabilizes neutrons, which rapidly decay into charged protons and electrons interacting with electromagnetic field (they were kept bound in the form of neutrons despite different size of protons and electrons only by the pressure of quantum degeneracy). Due to different masses of protons and electrons the net interaction with the electromagnetic field is non-zero despite equal opposite charges of both particles. The released electromagnetic  radiation in the region between photon sphere and Schwarzschild horizon accompanying the collapse of the initial Fermi sphere of neutrons converted into protons and electrons may be linked with some kind of frequently observed short giant gamma-ray cosmic bursts. An assessment of the energy accumulated in the initial Fermi sphere of neutrons in a neutron star at TOV-limit gives the energy of order of $10^{47}$ J (i.e., is equivalent to ca. Sun mass converted instantly into energy). This energy agrees with the energy of short isotropic giant gamma-ray bursts. The rest mass and angular momentum (including spin) of neutrons (decayed into protons and electrons and bereft of quantum statistics) are conserved in the final black hole core. The energy/mass conversion rate at the neutron Fermi sphere collapse  is here of ca. 30 \%. 

This rough scenario of the collapse of a neutron star merger with the mass exceeding TOV-limit needs more precise insight into the quantum statistics of identical indistinguishable particles and its flexibility to external topological factors including the specific trajectory homotopy near the gravitational singularity and the event horizon of the black hole. The definition of quantum statistics in terms of topological homotopy-type approach to systems of identical indistinguishable particles allows for discussion of quantum collective statistical oddness in the response to topological factors.  An illustration of such a flexibility of quantum statistics induced by the topological constraints is delivered in the case of electrons in regime of quantum Hall systems and has been convincingly supported by experimental observations of fractional quantum Hall effect occurring due to cyclotron topological modification of quantum statistics properties of particles in 3D spacetime (i.e., in the case of 2D spatial manifolds)     \cite{laughlin2,pra,annals2021}. In the present paper we propose to apply the homotopy topological approach to quantum statistics in 4D spacetime (3D spatial manifold in  gravitationally curved spacetime) for the  new concept of the rapid local decay of quantum statistics (and, in particular, local washing out of Pauli exclusion principle for particles bereft of their fermionic character) in the region of Schwarzschild zone below the innermost unstable circular orbit,  when a neutron star merger gains the mass above the TOV-limit.


\section{Homotopy of trajectories and Pauli theorem on spin and statistics}
\label{pauli}
Any discussion of quantum statistics must be  referred also to the commonly accepted theorem on the connection between spin and statistics formulated by Pauli \cite{pauli}.
This   theorem   states that quantum statistics of particles with half spin must be fermionic, whereas  of particles with integer spin -- bosonic. The conventional  proof of this theorem was supported by relativistic quantum arguments that for half spin Dirac electrodynamics, which admits Hamiltonian formulation for simultaneously particles and antiparticles,  the field operators  defining particles (or antiparticles) must  anticommute to assure the conservation  of time arrow (or, equivalently, positively defined kinetical energy and mass of free particles) \cite{landaushort,duck1}. This proof has been, however, limited to free particles without interaction, when the Hamiltonian is bilinear form with respect to field operators (annihilation and creation operators of second quantization). 
However, the Pauli theorem holds also for interacting particles and the conventional its proof reflects rather the fact that the Pauli theorem is robust against nonrelativistic-relativistic quantum transition only, and the proof is beyond the relativistic or nonrelativistic limits. In the latter case the linkage of half spin and fermionic statistics is deeply rooted as well as the linkage of integer spin with bosonic statistics. An exhausted discussion of the application of Pauli theorem on spin-statistics connection for nonrelativistic and relativistic particles, including the review of various trials of its  proof are presented in Refs \cite{duck1,duck2,balachandran}, supplemented  there also with a rigorous complete proof for noninteracting particles by Duck and Sudarshan \cite{duck1}. The Pauli  theorem on spin and statistics is true, however, independently of interaction. The actual proof of this theorem must be of topological type (robust to interaction) in view of homotopy based quantum statistics. Indeed, the Pauli theorem  follows directly from the coincidence of scalar unitary irreducible representations of the rotation group which define quantization of angular momentum and spin \cite{rumerfet}, with unitary representations of braid groups related to statistics \cite{mermin1979aaa}. This has been formerly noticed (in \cite{jac}, paragraph 3.2.5) that due to the overlap of some elements of the braid group and the rotation group  their unitary representations must  be linked as uniform on group generators. These representations are uniform on subsets collecting group generators, thus agree and mutually correspond if referred to full groups -- half spin representation of the rotation group  agrees  with the antisymmetric representation of the braid group in 3D, i.e., with $e^{i\pi}$ representation of generators of the permutation group, whereas the integer angular momentum representation of rotation group agrees with the symmetric $e^{i0}$ representation of the permutation group generators.  This is also supported by the extension of Pauli theorem onto 2D anyons with fractional statistics and continuous not quantized spin of 2D electrons  \cite{jac}.  In this case the scalar unitary  representations of the braid group generators are $e^{i\alpha},\;\alpha\in[0,2\pi)$, which agree with the scalar unitary representations of U(1) -- the  group related to commutating rotations in 2D space and thus to not quantized 2D spin -- of the form also of $e^{i\alpha}$ with continuous parameter $\alpha$. It means that $\alpha$-type anyons must have spin equal to $\alpha/2\pi$ \cite{jac}.   The topological proof of the Pauli theorem is immune to the particle interaction.

For 3D spatial manifolds in the case of particles in 4D spacetime, the covering group of the rotation group  $O(3)$ is $SU(2)$ of which irreducible representations fall into two classes characterizing integer and half-integer angular momenta. These two classes correspond with two distinct scalar unitary representations of the permutation group $S_N$, which is the full braid group for 3D manifolds. The representations of both groups coincide as they have some common elements.  For particles on 2D manifolds (3D spacetime) spin is not quantized as the $O(2)$ rotation group is isomorphic with $U(1)$ group possessing just the same unitary representations as the Artin group, the full braid group for $M=R^2$. In this case the Pauli theorem also holds though for not quantized spin and similarly continuously changing anyon statistics. 

The flexibility of quantum statistics with respect to some topological factors, which confine the availability of classical trajectories for particles, is independent of spin in fact. For 3D electrons maintaining the ordinary spin $\frac{1}{2}$ their dynamics can be ranged to 2D plane like in real experimental setups for quantum Hall effect (both integer or fractional \cite{klitzing1980,tsui1982}), where electrons are forced to move in thin layer with a few nanometre-depth. This layer is not strictly 2D, but electrons cannot hop one over another and have planar homotopy of trajectories though still have quantized 3D spin. Application of sufficiently strong perpendicular magnetic field can restrict braid trajectories (built of  cyclotron orbits of finite size in the plane topology) and can change quantum statistics not influencing, however,  spin of electrons \cite{laughlin2,pan2003,annals2021}. 

This indicates that statistics and spin, though coincide via the agreement between unitary representations of the rotation and the braid groups, are in fact independent, and one can imagine a situation when the spin is defined but the statistics not (the latter in the case of absence of the braid group). Such a situation happens in extremely strong gravitational field inside the black hole, beneath and close beyond the Schwarzschild horizon (beneath the innermost unstable circular orbit of a black hole),  as it will be discussed in the following paragraph.

	\section{Decay of quantum statistics beneath the innermost unstable circular orbit of a black hole}

A common feature of all singular solutions of Einstein general relativity equations is the unavoidable fall of mass and photons onto the central singularity if they found beneath the Schwarzschild surface. This one-way movement overwhelming the entire dynamics is conventionally associated with classical imagination of a black hole attracting and irrevocably capturing everything below the event horizon. This can be illustrated in the simplest Schwarzschild metric for isotropic spherically symmetrical massive body $M$ without rotation and electrical charge, which in spherical coordinates $(t,r,\theta,\phi)$ has the form for line element for proper time	\cite{schwarzschild},
\begin{equation}
	\label{metryka1}
	-c^2d\tau^2=-\left(1-\frac{r_s}{r}\right)	c^2dt^2+\left(1-\frac{r_s}{r}\right)^{-1}dr^2+r^2(d\theta^2+sin^2\theta d\phi^2),
\end{equation}
where $\tau$ is the proper time, $t$ is the time measured infinitely far of the massive body, $r_s=\frac{2GM}{c^2}$ is the Schwarzschild radius ($G$ is the gravitation constant, $c$ is the light velocity in the vacuum). The Schwarzschild metric has a singularity at $r=0$, which is an intrinsic curvature singularity. It also seems to have a singularity at the event horizon $r=r_{s}$. Depending on the point of view, the metric (\ref{metryka1}) is therefore defined only on the exterior region $r>r_{s}$ or on the interior region $r<r_{s}$. However, the metric is actually non-singular across the event horizon, as one sees in suitable coordinates. For $r\gg r_{s}$, the Schwarzschild metric is asymptotic to the standard Lorentz metric on Minkowski space. The Schwarzschild metric is a solution of Einstein field equations in empty space, meaning that it is valid only outside the gravitating body. That is, for a spherical body of radius $R$ the solution is valid for $ r>R$. To describe the gravitational field both inside and outside the gravitating body the Schwarzschild solution must be matched with some suitable interior solution at $r=R$ such as the interior Schwarzschild metric \cite{frolov}. In the case of classical concept of a black hole $R=0$ and the above described problem disappears. 

The singularity at $r = r_s$ divides the Schwarzschild coordinates in two disconnected patches. The exterior Schwarzschild solution with $r > r_s$ is the one that is related to the gravitational fields of stars and planets. The interior Schwarzschild solution with $0 \leq r < r_s$, which contains the singularity at r = 0, is completely separated from the outer patch by the singularity at $r = r_s$. The Schwarzschild coordinates therefore give no physical connection between the two patches, which may be viewed as separate solutions. The singularity at $r = r_s$ is an illusion however; it is an instance of what is called a coordinate singularity. As the name implies, the singularity arises from a bad choice of coordinates or coordinate conditions. When changing to a different coordinate system (for example Lemaitre coordinates, Eddington–Finkelstein coordinates, Kruskal–Szekeres coordinates, Novikov coordinates, or Gullstrand–Painlevé coordinates) the metric becomes regular at $r = r_s$ and can extend the external patch to values of $r$ smaller than $r_s$. Using a different coordinate transformation one can then relate the extended external patch to the inner patch.

The case $r = 0$ is different, however. If one asks that the solution be valid for all $r$ one runs into a true physical singularity, or gravitational singularity, at the origin. To see that this is a true singularity one must look at quantities that are independent of the choice of coordinates. One such important quantity is the Kretschmann invariant, which is given by,
\begin{equation}
	R^{\alpha\beta\gamma\delta}R_{\alpha\beta\gamma\delta}=\frac{12r_s^2}{r^6}=\frac{48G^2M^2}{c^4r^6},
\end{equation}
($R_{\alpha\beta\gamma\delta}$ is the Riemann curvature tensor) which shows that at $r= 0$ the curvature becomes infinite, indicating the presence of a singularity. It is 
a generic feature of the theory and not just an exotic special case.

For $r < r_s$ the Schwarzschild radial coordinate $r$ becomes timelike and the time coordinate $t$ becomes spacelike. A curve at constant $r$ is no longer a possible worldline of a particle, not even if a force is exerted to try to keep it there; this occurs because spacetime has been curved so much that the direction of cause and effect (the particle's future light cone) points into the singularity.

Hence, a particle placed beneath the Schwarzschild surface has a dynamics completely controlled by the central singularity and this particle tends to the singularity along the Schwarzschild geodesics. The equation for geodesics is as follows,
\begin{equation}
	\label{geodesics}
	\frac{d^2x^{\lambda}}{dq^2}+\Gamma^{\lambda}_{\mu\nu}\frac{dx^{\mu}}{dq}\frac{dx^{\nu}}{dq}=0,
\end{equation}
where variable $ q$ parametrizes the particle's path through spacetime, its so-called world line. The Christoffel symbols are defined by metric tensor $g_{\mu\nu}$ and for Schwarzschild metric attain the form, 
\begin{equation}
	\begin{array}{l}
		\Gamma_{rt}^t=-\Gamma^r_{rr}=\frac{r_s}{2r(r-r_s)},\;\Gamma_{tt}^r=\frac{r_s(r-r_s)}{2r^3},\;
		\Gamma_{\phi\phi}^r=(r_s-r)sin^2\theta,\;\\
		\Gamma_{\theta\theta}^r=r_s-r,\;\Gamma_{r\theta}^{\theta}=\Gamma_{r\phi}^{\phi}=\frac{1}{r},\;
		\Gamma_{\phi\phi}^{\theta}=-sin\theta cos\theta,\;\Gamma_{\theta\theta}^{\phi}=cot\theta,\\
	\end{array}
\end{equation}
which allows to solve Eq. (\ref{geodesics}) and determine geodesics (for detailed analysis of these geodesics cf. e.g., \cite{landaufield}), as briefly summarized in Appendix \ref{bbb}.

Particles falling on the singularity $r=0$ must travel along these one-way geodesics and no other trajectories exist for particles beneath the Schwarzschild surface. If one considers a collection of many particles ranged by the Schwarzschild radius then one can conclude that for these particles do not exist trajectories for their mutual exchanges
(as demonstrated in Appendix \ref{bbb}). Hence, the braid group for particles in this region cannot be defined as the generators of the braid group $\sigma_i$ (elementary exchanges of neighbouring particles at some their numeration, fixed but arbitrary, cf. e.g,, \cite{birman}) cannot be implemented. Thus, a quantum statistics for these particles loses locally its sense, it cannot be defined by scalar unitary representations of the braid group because of the absence of this group for the manifold confined to this specific range. Thus, the quantum statistics decays locally including also the local revoking of Pauli exclusion principle, as demonstrated in Appendix \ref{bbb}.

 The similar argumentation addressed to homotopy of particle trajectories precluding particle exchanges holds also in the close vicinity outside the Schwarzschild horizon, in the range encircled by the photon sphere with radius $3r_s/2$, which is simultaneously the limiting sphere with radius of the innermost unstable circular orbit  for arbitrary massive particles, cf. Appendix \ref{bbb}. At the photon sphere takes place the qualitative change of particle trajectory homotopy, which precludes particle exchanges below the innermost unstable circular orbit. This causes local disappearance of the braid group and thus local decay of quantum  statistics including local revoking of Pauli exclusion principle. This  does not mean a transition of fermions to bosons, which would violate the Pauli theorem. This theorem still holds but none quantum statistics is defined because of local absence of the braid group. Such a situation occurs in the case of neutron star when its Schwarzschild radius is comparable with the radius of the star, which happens above the TOV-limit. The disappearance of quantum statistics of identical particles in such a neutron star causes the instant its shrinking due to rapid relief in quantum degeneracy pressure and simultaneous the release of large number of photons emitted according to the Fermi golden rule scheme of quantum transitions, when neutrons stripped of statistics and thus decayed into charged protons and electrons, jump from the high energy stationary states in the initial Fermi sphere of fermions onto the ground state. Such a gamma-ray emission escaped from the region between photon sphere and Schwarzschild horizon can be responsible for some kind of frequently observed giant gamma radiation bursts in observable universe. 

\section{Linkage with short giant gamma-ray bursts}
In order to quantitatively estimate the energy released at the transformation of a cold neutron star into the black hole via proposed scheme of the decay of quantum statistics and local revoking of Pauli exclusion principle constraint, the energy of initial neutron Fermi system must be evaluated. For the idealized model we consider only neutrons neglecting other admixtures to real neutron star composition.  In a neutron star neutrons fill the Fermi sphere with a radius
\begin{equation}
	\label{promienf}
	p_F=\hbar (3\pi^2 n/V)^{1/3},
\end{equation}
where $n/V$ is the concentration of fermions -- $n$ neutrons in this case, in the volume $V$ of the star. The derivation of Eq. (\ref{promienf}) consists in application of the Bohr-Sommerfeld rule of quantization to the phase space volume $V\frac{4}{3}\pi p_F^3$, which corresponds to $n=2V\frac{4}{3}\pi p_F^3/h^3$  single-particle quantum states (where $h=2\pi \hbar$, $\hbar=1.05 10^{-34}$  J s is the Planck constant, additional factor 2 is due to doubling of states for particles with  spin $\frac{1}{2}$).  This Fermi radius is invariant to interaction of neutrons according to Luttinger theorem \cite{luttinger} and it is the same in noninteracting gaseous system and for arbitrarily strongly attracting or repulsing particles, provided that the concentration of particles $\frac{n}{V}$ is conserved. The Fermi energy in relativistic case is given by $\varepsilon_F=\sqrt{c^2 p_F^2+m^2 c^4}-mc^2$ with the neutron mass $m=1.675 \times 10^{-27}$ kg (in nonrelativistic case the Fermi energy was $\varepsilon_F=p_F^2/2m^*$, where an effective mass in the Fermi liquid equals to $m^*=m(1+F_1)$ and $F_1$ is the first Landau amplitude to account interaction of fermions in isotropic system, cf. e.g., \cite{abrikosov1975}). For rough estimation we use the model of free fermions with momentum as a good quantum number (as for locally translationally invariant systems) to estimate an order of the magnitude of considered quantities. 
The density of a neutron star, i.e., its mass per volume, accounts for the rest mass of neutrons (in an idealized model of neutron star composition) and the mass equivalent to the total energy of the Fermi sphere for neutrons and their interaction (the latter interaction neglected here, similar as for nonrelativistic case with $F_1=0$). 
If the density of a star is denoted by $\rho$, the number of all neutrons in the star by $ n$, and the radius of the initial star by $r$, then 
\begin{equation}
	\label{system}
	\begin{array}{l}
		\rho \frac{4}{3}\pi r^3 = m n +E/c^2,\\
		E=\int_0^{p_F}dp \int_0^{2\pi} d\phi\int_0^{\pi} d\theta \varepsilon(\mathbf{p}) p^2 sin\theta\frac{1}{(2\pi \hbar)^3}\times \frac{4}{3}\pi r^3,\\
		p_F= \hbar (3\pi^2 n/(\frac{4}{3}\pi r^3))^{1/3},\\
	\end{array}
\end{equation}
where $m$ is the bare rest mass of the neutron ($m=1.675 \times 10^{-27}$ kg).
The above system of equations can be solved for initial values of $\rho$ and $r$ of the neutron star, giving the Fermi radius $p_F$ and the total energy $E$ of the Fermi sphere of neutrons in the whole star. Some examples of solution of the system of equations (\ref{system}) for few selected $\rho$ and $r$ are listed in Tables \ref{tabue}, \ref{tabueee} and \ref{tabueee} for three various definitions of neutron kinetical energy $\varepsilon(\mathbf{p})$, for comparison.

\begin{table}
	\centering
	\begin{tabular}{|l|l|l|l|l|l|}
		\hline
		$\rho$ [kg/m$^3$] & $r$ [km] & $n$& $p_F$ [kg m/s]& $\varepsilon_F$ [GeV] & $E$ [J] \\
		\hline
		$5\times10^{18}$& $10$ & $1.16\times 10^{58}$&$4.5\times 10^{-19}$ &$0.37$ & $2 \times 10^{47}$ \\
		\hline
		$2 \times 10^{18}$&$10$& $4.7 \times 10^{57}$&$3.37 \times 10^{-19}$&$0.2$ &$4.8 \times 10^{46}$\\
		\hline
		$1.0 \times 10^{19}$& $8$&$1.08 \times 10^{58}$&$5.57\times 10^{-19}$&$0.58$&$3 \times 10^{47}$\\
		\hline
		$2.5 \times 10^{18}$&$8$&$2.97 \times 10^{57}$&$3.62 \times 10^{-19}$&$0.24$&$3.5 \times 10^{46}$\\
		\hline
	\end{tabular}
	\caption{Fermi momentum $p_F$, Fermi energy $\varepsilon_F$ (for nonrelativistic case of estimation, $\varepsilon(\mathbf{p})=p^2/2m$, for comparison only as in this case we deal rather with relativistic case in almost whole Fermi sphere) and total energy of the Fermi sphere $E$ released at the decay of the statistics for selected examples of density $\rho$ and radius $r$ of a neutron star and found via solution of equation system (\ref{system}) ($n$ total number of neutrons in the star). }
	\label{tabue}
\end{table}

\begin{table}
	\centering
		\begin{tabular}{|l|l|l|l|l|l|}
		\hline
		$\rho$ [kg/m$^3$] & $r$ [km] & $n$& $p_F$ [kg m/s]& $\varepsilon_F$ [GeV] & $E$ [J] \\
		\hline
		$5\times10^{18}$& $10$ & $1.13\times 10^{58}$&$4.52 \times 10^{-19}$ &$0.32$ & $1.84 \times 10^{47}$ \\
		\hline
		$2 \times 10^{18}$&$10$& $4.70 \times 10^{57}$&$3.38 \times 10^{-19}$&$0.19$ &$4.47 \times 10^{46}$\\
		\hline
		$1.0 \times 10^{19}$& $8$&$1.02\times 10^{58}$&$5.61\times 10^{-19}$&$0.47$&$2.69 \times 10^{47}$\\
		\hline
		$2.5 \times 10^{18}$&$8$&$2.99 \times 10^{57}$&$3.63 \times 10^{-19}$&$0.22$&$3.24 \times 10^{46}$\\
		\hline
	\end{tabular}
	\caption{Fermi momentum $p_F$, Fermi energy $\varepsilon_F$ (for relativistic case of estimation, $\varepsilon(\mathbf{p})=\sqrt{c^2p^2+m^2c^4}-mc^2$) and total energy of the Fermi sphere $E$ released at the decay of the statistics for selected examples of density $\rho$ and radius $r$ of a neutron star and found via solution of equation system (\ref{system}) ($n$ total number of neutrons in the star). }
	\label{tabuee}
\end{table}

\begin{table}
	\centering
		\begin{tabular}{|l|l|l|l|l|l|}
		\hline
		$\rho$ [kg/m$^3$] & $r$ [km] & $n$& $p_F$ [kg m/s]& $\varepsilon_F$ [GeV] & $E$ [J] \\
		\hline
		$5\times10^{18}$& $10$ & $9.5\times 10^{57}$&$3.21\times 10^{-19}$ &$0.60$ & $4.55 \times 10^{47}$ \\
		\hline
		$2 \times 10^{18}$&$10$& $4.03 \times 10^{57}$&$3.37 \times 10^{-19}$&$0.63$ &$1.46 \times 10^{47}$\\
		\hline
		$1.0 \times 10^{19}$& $8$&$4.46 \times 10^{57}$&$4.9\times 10^{-19}$&$0.92$&$8.06 \times 10^{47}$\\
		\hline
		$2.5 \times 10^{18}$&$8$&$2.17 \times 10^{57}$&$3.26 \times 10^{-19}$&$0.61$&$1.05 \times 10^{47}$\\
		\hline
	\end{tabular}
	\caption{Fermi momentum $p_F$, Fermi energy $\varepsilon_F$ (for ultrarelativistic case of estimation, $\varepsilon(\mathbf{p})=c p$) and total energy of the Fermi sphere $E$ released at the decay of the statistics for selected examples of density $\rho$ and radius $r$ of a neutron star and found via solution of equation system (\ref{system}) ($n$ total number of neutrons in the star). }
	\label{tabueee}
\end{table}

For a density of a neutron star merger of order of $10^{18-19}$ kg/m$^3$ the number of neutrons including contribution to the star mass of total energy of the Fermi sphere, $n \sim 10^{58}$ (cf. first equation in (\ref{system})). For such a concentration of fermions, $p_F \sim 10^{-19}$ kg m/s and the Fermi energy of neutrons, $\varepsilon_F =\varepsilon(p_F) \sim 0.2 - 0.6 $ GeV. The Fermi energy is the uppermost energy of neutrons in the star and concerns only highest in energy  neutrons on the surface of the Fermi sphere. Other neutrons (majority of them) have lower energies determined by their positions in the Fermi sphere.  The total energy of the Fermi sphere of neutrons in the whole neutron star  can be estimated as,
\begin{equation} 
	E=\int_0^{p_F}dp \int_0^{2\pi} d\phi\int_0^{\pi} d\theta \varepsilon(\mathbf{p}) p^2 sin\theta\frac{1}{(2\pi \hbar)^3}\times \frac{4}{3}\pi r^3,
\end{equation}
where $r$ is the radius of the neutron star and $\varepsilon(\mathbf{p})=\frac{p^2}{2m}$, $\sqrt{c^2 p^2+m^2c^4}-mc^2$, $ cp$ is the energy of the neutron with momentum $ \mathbf{p} $ for nonrelativistic (Table \ref{tabue}), relativistic (Table \ref{tabuee}) and ultrarelativisic (Table \ref{tabueee}) case, respectively.

\begin{figure}
	\centering
	\includegraphics[width=0.65\columnwidth]{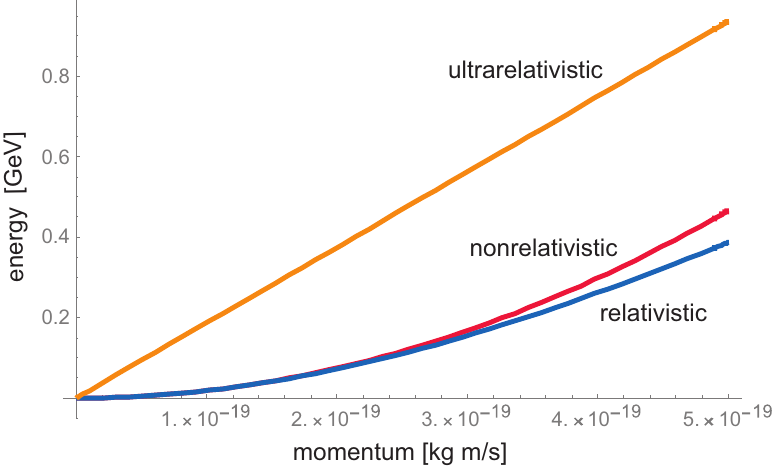}
	\caption{\label{222} Comparison of neutron energies in relativistic ($\sqrt{c^2 p^2+ m^2 c^4}-mc^2$), nonrelativistic ($p^2 /2m$) and ultrarelativistic approximation ($cp$) in the range of its momentum in a neutron star.	}
\end{figure} 

For exemplary radius of the neutron star $r=8 -10 $ km the energy release, when all neutrons from the Fermi sphere fall onto zeroth energy ground state, is thus of order of $E\sim 0.5\times 10^{47} -3 \times 10^{47}$ J, which is of the same order as the energy equivalent to the Sun mass ($M_{\odot}c^2$, which for the Sun mass $M_{\odot}=1.99 \times 10^{30}$ kg gives $1.8 \times 10^{47}$ J). This is just as estimated averaged energy of observed short giant gamma-ray cosmic bursts (assuming the isotropic radiation from their sources). 

The remnants compressed into the quantum 'singularity' of the arisen black hole still have the rest mass $n m$ of all neutrons, and for four examples from Table \ref{tabue} are of the mass $1.9 \times 10^{31}$ kg, $7.8 \times 10^{30}$ kg, $1.8 \times 10^{31}$ kg and $4.97 \times 10^{30}$ kg, respectively. For relativistic and nonrelativistic case the energy of Fermi sphere to be released due to decay of statistics is similar, as is illustrated in Tables \ref{tabuee} and \ref{tabue}. This similarity in estimation follows from the fact that in the range of momentum of neutrons in a neutron star, the relativistic kinetical energy $\varepsilon(\mathbf{p})=\sqrt{c^2p^2+m^2c^4}-mc^2$ is not much different than nonrelativistic $\varepsilon(\mathbf{p})=\frac{p^2}{2m}$, cf. Fig. \ref{222}. For ultrarelativistic case with $\varepsilon(\mathbf{p})=cp$ the energy estimation is higher by a factor of ca. 2 (Table \ref{tabueee}), but this ultrarelativistic limit does not hold for neutrons in a neutron star (it requires $cp\gg mc^2$, which is not satisfied for neutrons in the range of their momentum in a neutron star, as for neutrons $mc^2\simeq 0.9 $ GeV). 

The initial energy of the Fermi sphere of neutrons can be released during the collapse of this Fermi sphere in the form of gamma-ray photons. Neutrons not blocked by Pauli exclusion principle rapidly and coherently decay into charged components, protons and electrons, with different masses and thus with large net coupling to electromagnetic field. Despite balanced opposite charges the mass difference causes nonzero coupling to electromagnetic field in the scheme of quantum transitions according to the Fermi golden rule.

At the collapse of the neutron star merger exceeding in mass the TOV-limit, the radiation release at the decay of Fermi sphere of neutrons can escape only from the region between the photon sphere and Schwarzschild horizon if fact, whereas the rest beneath the horizon is unavoidably captured by the central singularity. Form point of view of the remote observer, the ratio of the volume of the photon zone (the region between the photon sphere with  radius of the innermost unstable circular orbit $r=1.5 r_s$, cf. Appendix \ref{bbb}) to the Schwarzschild horizon sphere is ca $2.4$. This means that in the above estimations at most ca. $2/3$ of the radiation released at neutron Fermi sphere collapse in the whole volume of the merger can be observable as external radiation, which, however, does not change the magnitude orders in the above estimations.

\section{Linkage with luminosity of quasars}

Let us consider the exemplary  quasar with the central giant black hole  consuming 10 $M_{\odot}$ per year of the hydrogen gas from the accretion disc, i.e., ca.  $0.1$ Earth mass per second. Let us assume the stable uniform in time process of matter accretion. The transport of  the matter across the accretion  disk is steady for a relatively long time,  thus one can perform the calculation of the mass to energy conversion rate per finite time period, let say  per   the single second. The gas in accretion disc is ionized due to friction and then step by step compressed when is approaching the Schwarzschild surface of the central black hole. At sufficiently high concentration of electrons and protons (the same for both types of fermions due to charge neutrality of the plasma) both their Fermi liquids are quantumly degenerated, i.e., their chemical potentials  exceed
the local temperature ($\mu_e,\mu_p\gg k_BT$, where $\mu_{e(p)}$ is the chemical potential of electrons (protons) and $k_B=1.38 \times 10^{-23}$ J/K, $T$ is the local temperature in the accretion disc). For a local concentration of fermions (electrons or protons) $\rho(r)=\frac{dn}{dV(r)}=\frac{N}{V(r)}$ (where $dn$ is the portion of electrons (or protons) infalling in the accretion disc the region with the radius {r} in time $dt$ and compressed to the volume $dV(r)$ depending on the distance form central gravitational singularity denoted here by $r$. Due to the stationary conditions this local concentration is equal to the total number of electrons (or protons) $N$ per second  compressed locally to $V(r)$) one can find the Fermi momentum of electrons (exactly the same for protons in the  neutral plasma),
\begin{equation}
	\label{fm}
	p_F(r)=\hbar\left(    3\pi^2 \rho(r)\right)^{1/3}=\hbar\left(3 \pi^2 \frac{N}{V(r)}\right)^{1/3}.
\end{equation}
Incoming  portions of electrons and protons in radial direction towards the central singularity  add up in due of one second time period to the total constant flow of mass (in the example, of $0.1 \times M_Z$ kg/s, the Earth mass $M_Z=5.97 \times 10^{24}$ kg). The locally accumulated energy in the Fermi spheres of electrons and protons grows with lowering $r$ due to the increase of the compression caused by the gravitation field (i.e., due to the lowering of $V(r)$). This energy is proportional to $V(r)$, which influences moreover the Fermi momentum $p_F(r)$ via Eq. (\ref{fm}),  and the energy equals to
\begin{equation}
	\label{enn}
	\begin{array}{l}
		E(r)=E_e(r)+E_p(r),\\
		E_e(r)=\frac{V(r)}{2 \pi^2 \hbar^3}\int_o^{p_F(r)}dp p^2 \left( \sqrt{p^2 c^2 +m_e^2 c^4}-m_e c^2\right),\\ 
		E_p(r)=\frac{V(r)}{2 \pi^2 \hbar^3}\int_o^{p_F(r)}dp p^2 \left(\sqrt{p^2 c^2 +m_p^2 c^4}-m_p c^2\right),\\
	\end{array}
\end{equation}
where the energy $E_{e(p)}$ refers to electrons (protons).

At  some critical radius $r^*$ close to Schwarzschild zone (cf. Appendix \ref{bbb} where we demonstrate that $r^*=1.5r_s$, i.e., is the radius of the innermost unstable circular orbit, which is simultaneously the radius of the photon sphere in the case of non-rotating and non-charged black hole) the decay of  quantum statistics takes place due to  topology reason (cf. Appendices \ref{aaa} and \ref{bbb}) and both Fermi spheres of electrons and protons collapse. This undergoes by portions $dn$ of particle flow incoming to $r^*$ region in infinitely small time periods $dt$, which add up  to  $0.1 \times  M_Z$ kg per second. The value  of the energy released depends on the Fermi momentum at $r=r^*$ and  attains  $10^{40}$ J at sufficiently high level of compression, i.e., at sufficiently small $V(r^*)$ determined from the self-consistent system of Eqs (\ref{fm}) and (\ref{enn}).

Note that if $V$ in Eq. (\ref{fm}) is treated as the relativistic proper volume, then outside the Schwarzschild zone the proper volume element scales as (in the Schwarzschild metric (\ref{metryka1})),
	\begin{equation}
		\label{correction}
		\left(1-\frac{r_s}{r}\right)^{-1/2}dr r^2 sin\theta d\theta d\phi.
	\end{equation}
	At $r=r^*=1.5r_s$ the factor renormalizing the same volume observed by a remote observer, $dr r^2 sin\theta d\theta d\phi$, is ca. 1.7, which gives the reduction  of $p_F$ (in Eq. {\ref{fm}}) caused by gravitation curvature by factor ca. $1.7^{1/3}\simeq0.84$, which does not change orders in the above estimations. The change of $p_F$ by one order of the magnitude would need the closer approaching the Schwarzschild horizon, at $r\simeq 1.000001 r_s$, i.e., rather distant from the critical $r^*=1.5 r_s$. Hence, for the rough estimation of the effect of Fermi sphere collapse  the correction (\ref{correction})
	is unimportant and can be included as the factor $0.84$ to right-hand side of Eq. (\ref{fm}), which does not change the orders in the energy estimation (in Eq. (\ref{enn})).

To the initial mass of the accreted  gas (assuming to be composed of hydrogen $H$)
contribute mostly protons (ca. 2000 times more massive than electrons), thus the total number of electrons per one second, the same as the number of protons, equals to, $N\simeq 0.1 M_Z /m_p\simeq 3.57 \times 10^{50}$. Solving simultaneously Eqs (\ref{fm}) and (\ref{enn}), assuming $N=3.57 \times  10^{50}$  in volume $V(r^*)$ and released energy $E(r^*)=10^{40}$ J,  we find the volume (per second) of compressed plasma $V(r^*)=0.8 \times 10^5$ m$^3$ and electron or proton Fermi sphere radius $p_F(r^*)=5.3 \times 10^{-19}$ kg m/s. Electrons and protons are compressed to the same volume $V(r^*)$ (due to neutrality of plasma), hence,  their  concentration at $r^*$, $\rho(r^*)=4.45 \times 10^{45}$ 1/m$^3$. The mass density at $r^*$ (including mass equivalent to the energy stored up in Fermi spheres of electrons and protons) is thus  $\xi(r^*)=\frac{0.1 M_Z}{V(r^*)}+\frac{E(r^*)}{c^2 V(r^*)}\simeq 9 \times 10^{18}$ kg/m$^3$, or with the correction (\ref{correction}), $5.3 \times 10^{18}$ kg/m$^3$ (similar to mass density in neutron stars). The  released energy of $E(r^*)=10^{40}$ J is equivalent to 30 \% of the infalling mass of $0.1$ Earth mass (per second). It means that the compressed plasma with degenerate Fermi liquid of electrons (and also of protons) is by 30 \% more massive than initial remote diluted gas. This increase of the mass is caused by the gravitational field of the central black hole, which compresses both  systems of fermions and raises the energy stored up in Fermi spheres of fermions. The energy of  gravitation field is accumulated in local Fermi spheres of electrons and protons.  The ratio of total Fermi sphere energies of electrons and protons is $\frac{E_n(r^*)}{E_p(r^*)}\simeq 1.4$. The Fermi energy (the uppermost energy in the Fermi sphere) of electrons with Fermi momentum $p_F=5.48 \times  10^{-19}$ equals to $\varepsilon_F= 0.6 $ GeV (it is the upper possible energy of emitted photons), which in thermal scale (in units of $k_B=1$) is of order of $7.5 \times 10^{12}$ K -- this makes that electron liquid is quantumly degenerated at lower temperatures (quasars are not source of thermal gamma radiation, thus their actual temperatures are much lower). The Fermi energy of protons with Fermi momentum $p_F=5.48 \times  10^{-19}$ equals to $\varepsilon_F= 0.2 $ GeV (it is the upper possible energy of emitted photons by jumping of protons), which in thermal scale (in units of $k_B=1$) is of order of $2.1 \times 10^{12}$ K. 

The release  of the energy  due to the  collapse of the  Fermi sphere of charged particles
undergoes according to the quasiclassical Fermi golden rule scheme for quantum transitions \cite{landau1972}, when such transitions are admitted by the local revoking of Pauli exclusion principle (cf. Appendix \ref{bbb}). Charged carries (electrons and protons) couple to electromagnetic field and the matrix element of this coupling between individual particle state in the Fermi sphere and its ground state  is the kernel of the Fermi golden rule formula for transition  probability per time unit, $\delta w_{i\rightarrow f}$, for this particle,
\begin{equation}
	\delta w_{i\rightarrow f}= \frac {2\pi}{h}\left|<i| \hat{V}|f>\right|^2\delta(\varepsilon_i-\varepsilon_f-\hbar\omega),
	\end{equation}
where $<i|$ and $<f|$ are quantum states of a fermion in the Fermi sphere, initial and final ones, respectively, $\varepsilon_i$ and $\varepsilon_f$ are energies of these states and $\hbar \omega$ is the energy of emitted photon. The Dirac delta $\delta(\varepsilon_i-\varepsilon_f-\hbar\omega)$ assures energy conservation at quantum transition $i\rightarrow f$ assisted by photon emission.   
 The interaction of charged particles with photons in above formula, $\hat{V}$, is proportional to the electromagnetic field strength, thus the increasing number of excited photons strengthens the coupling in the similar manner as at stimulated emission (known from e.g.,  laser action) and accelerates the quantum transition of the Fermi sphere collapse. 

We see that the same quantum  mechanism can be responsible for short giant gamma-ray bursts accompanying the rapid merging of neutron stars eventually converted into a black holes and the enormously high and  stable for a long time luminosity of quasars. In both cases the rate of mass to energy  conversion is of order of 30 \%, unreachable in any other known process except for matter-antimatter annihilation. In the case of quasars  the described above mechanism of mass to energy conversion near the  black hole event horizon  supplements the conventional  models of the accretion disc luminosity from more distant regions of the disc based on the classical hydrodynamical approach to matter accretion  \cite{sunaev,merloni,novikov,abramowiczg}. The region closest to the Schwarzschild horizon  is below the inner edge of the disc assumed in conventional hydrodynamical models, but this region appears highly significant in view of the presented above quantum-type large contribution to the total quasar luminosity.

\section{Conclusions}
We have proposed a quantum mechanism of transformation of gravitation field energy into electromagnetic radiation in close vicinity of Schwarzschild horizon of a black hole. This mechanism manifests itself at the  collapse into a black hole of a neutron star merger with the mass exceeding the TOV-limit. By analysis of particle trajectories beneath the Schwarzschild surface and beneath the photon sphere (coinciding with the sphere with radius of the innermost unstable circular orbit for massive particles) we have concluded that the quantum statistics cannot be assigned to particles in these regions, because of specific homotopy of admissible particle trajectories here, which locally precludes the braid group definition. The related local decay of quantum statistics, because of absence of the braid group, is analogous here to the flexibility and variation of quantum statistics versus a topological factors in the well developed experimental domain of fractional quantum Hall effect, where the change of topology of classical trajectories causes qualitative changes of corresponding quantum statistics defined by scalar unitary representations of the appropriate  braid groups. In the case of an unavoidable one-way falling of particles onto the central singularity of a black hole (described classically in terms of the Schwarzschild metric as one-way spirals onto the event horizon and the central singularity, the latter in terms of Kruskal-Szekeres or Novikov metrics) any braid group cannot be implemented for identical indistinguishable particles beneath the Schwarzschild horizon, as no trajectories for particle exchanges exist there. Similar homotopy of trajectories precludes particle exchanges  below the photon sphere being the same as the sphere with the radius of innermost unstable circular orbit for any massive particles (i.e., for $r<\frac{3r_s}{2}$, $r_s$ is the Schwarzschild radius). When the quantum statistics cannot be defined in these regions, the Pauli exclusion principle is locally revoked. This does not violate the Pauli theorem on spin and statistics, which still holds but is effective only if both  unitary representations of the rotation group and the braid group exist and  coincide. Beneath and closely beyond the Schwarzschild horizon up to the innermost unstable circular orbit,  the braid group cannot be, however, defined, though spin still maintains its significance. The unique situation of the overwhelming the whole dynamics by the central gravitational singularity of the black hole manifest thus itself in the instant local decay of the quantum statistics of indistinguishable particles falling onto this singularity. The homotopy of classical trajectories in the Schwarzschild region up to the innermost unstable circular orbit precludes particle exchanges and dismisses quantum statistics definition. This can happen when a neutron star merger with growing mass shrinks to the size comparable with its Schwarzschild sphere (the Schwarzschild radius grows proportionally to the mass, whereas the neutron star shrinks with increasing mass). 

The rapid reduction of the quantum degeneracy repulsion of fermions reliefs the internal pressure and the neutron star merger collapses into the black hole. This rapid quantum phase transition is accompanied by the release of photons emitted according to the Fermi golden rule for quantum transitions during falling of neutrons (coherently decayed into protons and electrons as being unstable at the absence of Pauli exclusion principle constraint) from their positions in the Fermi sphere onto their ground state. The estimation of Fermi energy (uppermost energy of fermions on  the Fermi surface) of neutrons in neutron star with the density ca. $10^{18-19}$ kg/m$^3$ is of order of GeV. For a neutron star with the radius of ca. $8-10$ km and the density of order of $10^{18}$ kg/m$^3$, the total energy accumulated in the Fermi sphere and released in the form of  emitted photons is ca. $10^{47}$ J during the Fermi sphere collapse, which agrees with short giant gamma-ray bursts observed averagely one per day with just this energy (assuming isotropic radiation of a source, the gamma-ray bursts have the energy just of order of $10^{47}$ J).

The same mechanism can contribute also to another high energy astrophysical process observed in quasars with their giant luminosity of order of $10^{40}$ W. Quasar with super-massive galactic central black hole consumes ionized gas from the accretion disc with the typical rate of order of the 10 Sun mass per year  ($0.1$ mass of the Earth per second). The mass to energy conversion ratio must be of order of ca. 30 \% to continuously  produce the luminosity of $10^{40}$ W for a long time. Such an effectiveness has the decay of Fermi spheres of electrons and protons in  highly compressed plasma in the accretion disc close to the Schwarzschild horizon (at the critical distance equal to the innermost unstable circular orbit of the central supermassive black hole). The energy accumulated in these Fermi spheres due to compression caused by the gravitation singularity and next released at Fermi sphere collapse, estimated for infalling of $0.1$ Earth mass per second is just of order of $10^{40}$ J per second, which agrees with quasar luminosity of $10^{40}$ W.

The collapse of the Fermi sphere of fermions near the Schwarzschild horizon of a black hole  is highly efficient,  with ca. 30 \% of energy/mass conversion  ratio and can help to elucidate the high energy astrophysical phenomena like some kind of giant gamma-ray bursts and visible luminosity of remote quasars, but holds for any black hole provided that the infalling matter is compressed at the innermost unstable circular orbit to the limiting density similar to that in neutron stars. No other so effective physical mechanism of mass to energy conversion is known except of matter-antimatter annihilation.

\appendix

	\section{Why scalar unitary representations of braid groups define quantum statistics}

	\label{aaa}
	The quantum indistinguishability of identical particles leads to quantum statistics of fermionic or bosonic type for particles in 4D spacetime, which differ by antisymmetry or symmetry of their multiparticle wave functions at exchanges of arguments in these functions. This is conventionally addressed to commutation or anticommutation of related local field operators of creation and annihilation of these particles. However, the quantum statistics is a collective non-local effect with topological roots (cf. \cite{leinaas1977} for first observations). A modern way to the formal definition of quantum statistics in multiparticle systems is offered by the quantization method upon Feynman path integration scheme. Though originally defined by Feynman for a single particle \cite{feynman1964}, the path integral can be generalized for $N$ identical indistinguishable particles which classically have trajectories in the multiparticle configuration space $F_N=(M^N-\Delta)/S_N$, as defined in paragraph \ref{uuu} (i.e., $M$ is a manifold on which $N$ particles are distributed, $M^N=M\times M\times\dots\times M$ is $N$-fold product of the manifold $M$, $\Delta$ is the diagonal subset of $M^N$ and $S_N$ is the permutation group of $N$ elements). The space $F_N$ is usually multiply connected, i.e., the fundamental group of $F_N$ -- the first homotopy group $\pi_1(F_N)$ is nontrivial \cite{mermin1979aaa,spanier1966}.  $\pi_1(F_N)$ is called as the full braid group of the system of $N$ identical and indistinguishable particles located on the manifold $M$ \cite{birman,mermin1979aaa} and trajectories (including closed loops which build $\pi_1(F_N)$) are $N$-strand bunches. 
	
\begin{figure}
		\centering
		\includegraphics[width=0.65\columnwidth]{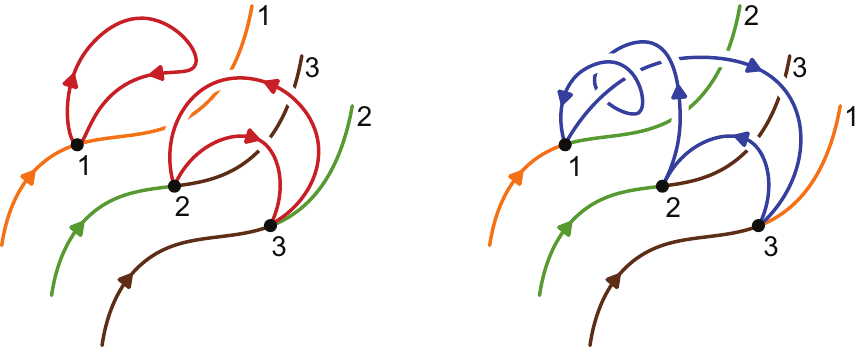}
		\caption{\label{gow11111} To a multiparticle trajectory in the configuration space (in the illustration for $N=3$, the configuration space of indistinguishable identical particles is $F_3=(M^3-\Delta)/S_3$) one can add an arbitrary loop from the full braid group $\pi_1(F_3)$ -- indicated as red and blue lines for  simple examples. Due to nonhomotopy of various braids from the full braid group (linking positions of particles at same intermediate time instant that may differ by a permutation) the trajectories with various braids attached are topologically nonequivalent, i.e., cannot be transformed one into another by continuous deformations -- they are nonhomotopic.	}
	\end{figure}

	Scalar unitary representations (1DURs, one dimensional unitary representations) of the braid group define quantum statistics of particles on the manifold $M$. The linkage od 1DURs of braid groups with quantum statistics can be reasoning by quantization within Feynman path integral formulation. The Feynman path integral defines the quantum propagator, i.e., the matrix element of the quantum evolution operator in the position representation, and for a single particle it has the form \cite{feynman1964}, 
	\begin{equation}
		\begin{array}{ll}
			I(\mathbf{r}, t; \mathbf{r}', t')=
			\int d\lambda e^{iS[\lambda(\mathbf{r}, t; \mathbf{r}', t')]/\hbar},
		\end{array}
		\label{path0}
	\end{equation}
	where $S[\lambda(\mathbf{r}, t; \mathbf{r}', t')]$ is the classical action of the particle along the trajectory $\lambda$ starting from point $\mathbf{r}$ at time instant $t$ and finishing in $\mathbf{r}'$ at time instant $t'$. Summation (integration with the measure $d\lambda$ in the trajectory space)
	of contributions of all possible trajectories joining fixed start and final points gives the complex amplitude 
	$I(\mathbf{r}, t; \mathbf{r}', t')$ of the probability for quantum transition between these points, i.e., the matrix element of the evolution operator in position representation.
	One can generalize the path integral (\ref{path0})
	onto the case of the system of $N$ identical indistinguishable particles with the multidimensional coordinate space $F_N=(M^N-\Delta)/S_N$. 
	For such $N$-particle system the path integral attains the form \cite{wu,lwitt,chaichian2},
	\begin{equation}
		\begin{array}{l}
			I(\mathbf{r}_1,\dots, \mathbf{r}_N, t; \mathbf{r}'_1,\dots, \mathbf{r}'_N, t')\\
			=\sum_{l\in \pi_1(F_N)}e^{i\alpha_l}\int d\lambda_l e^{iS[\lambda_l(\mathbf{r}_1,\dots, \mathbf{r}_N, t; \mathbf{r}'_1,\dots, \mathbf{r}'_N, t')]/\hbar},\\
		\end{array}
		\label{path}
	\end{equation}
	where $I(\mathbf{r}_1,\dots, \mathbf{r}_N, t; \mathbf{r}'_1,\dots, \mathbf{r}'_N, t')$ is the propagator, i.e., the matrix element of the quantum evolution operator of the total $N$-particle system in the position representation that determines the probability amplitude (complex one) for quantum transition from the initial point $(\mathbf{r}_1,\dots, \mathbf{r}_N)$ in the multiparticle coordination space $F_N$ at time instant $t$ to the other point $(\mathbf{r}'_1,\dots, \mathbf{r}'_N)$ in the space $F_N$ at time instant $t'$. However, due to particle indistinguishability the numeration of particles can be arbitrarily changed during the way between distinct points in $F_N$, which can be imagined by adding an arbitrary loop -- a braid from the full braid group $\pi_1(F_N)$ -- to arbitrary point of $N$-strand open trajectory in $F_N$ (cf. simple illustration in Fig. \ref{gow11111}).
	Because of impossibility to transform continuously various loops from the braid group one into other, the whole space of paths of $N$ particles (the domain of the path integral)	decomposes into separate disjoint subdomains numbered by elements of the braid group. Hence, it is impossible to define an uniform measure for path integration over the whole domain (due to discontinuity between its sectors) and 
	in Eq. (\ref{path}) $d\lambda_l$ is a separate measure in the path space sector numbered by the $l$-th element of the braid group $\pi_1(F_N)$ (braid groups are always countable or finite because they are generated by the finite number of generators, cf. e.g., \cite{birman}). $S[\lambda_l(\mathbf{r}_1,\dots, \mathbf{r}_N, t; \mathbf{r}'_1,\dots, \mathbf{r}'_N, t')]$ is the classical action for the trajectory $\lambda_l$ joining selected points in the configuration space $F_N$ between time instances $t$, $t'$ and lying in $l$-th sector of the trajectory space with the $l$-th braid loop attached to multi-strand open trajectories. The contributions of all disjoint trajectory domain sectors numbered by the braid group element discrete index $l$ (as for the countable group) must be summed together in final path integral. The components must enter the total sum (\ref{path}) with some arbitrary in general but unitary weight factor $e^{i\alpha_l}$ (unitary follows from the causality constraint in quantum evolution). It has been proved \cite{lwitt} that the unitary factors (the weights) associated with the contributions of disjoint sectors of the path integral domain, $e^{i\alpha_l}$ in Eq. ({\ref{path}), establish a one-dimensional unitary representation (1DUR) of the full braid group. Distinct unitary weights in the path integral (i.e., distinct 1DURs of the braid group) determine different types of quantum particles corresponding to the same classical ones. Braids describe particle exchanges (they are closed loops in $F_N$ with unified points distinct only in particle numeration), and thus their 1DURs assign quantum statistics in the system.
		
		It has been mathematically proved (e.g., in \cite{birman}) that for $N\geq2$ and $dim M\geq3$ (4D spacetime belongs to this class) the braid group $\pi_1(F_N)=S_N$, i.e., it is the finite permutation group. Any permutation group has only two distinct 1DURs, 
		\begin{equation}
			\sigma_i\rightarrow \left\{ 
			\begin{array}{l} 
				e^{i0},\\
				e^{i\pi},\\
			\end{array} \right.
		\end{equation}
		defining either bosons (for $e^{i0}$) or fermions (for $e^{i\pi}$). In the above formula $\sigma_i$ with $i=1,\dots,N-1$ are the generators of the braid group (in 4D spacetime, the permutation group), which are the elementary exchanges of positions of $i$-th particle with $(i+1)$-th one, at arbitrary but fixed numeration of all $N$ particles in the system. For 4D spacetime (or in higher dimensions) $\sigma_i^2=e$ (neutral element) and therefore the braid groups at such space dimensions coincide with the permutation group with only two distinct 1DURs. 
		
		For planar manifolds in the case of 3D spacetime, like for $M=R^2$ (or locally planar as the surface of a 3D sphere or torus), the braid group is not equal to $S_N$ and instead is infinite highly complicated countable group \cite{mermin1979aaa,birman}. Braid groups are always multi-cyclic groups (thus countable) generated by $\sigma_i$ generators -- exchanges of neighbouring particles (at some numeration of all particles), but for $M=R^2$ $\sigma_i^2\neq e$. For $M=R^2$ the braid group has been originally described by Artin \cite{artin1947}. Artin group has infinite number of 1DURs, $\sigma_i\rightarrow e^{i\alpha}$, $\alpha \in[0,2\pi)$, and the related distinct quantum statistics corresponding to various $\alpha$ can be associated with so-called anyons (the name was coined by Wilczek \cite{wilczek} and addressed to possible fractional statistics beyond fermionic or bosonic ones). 
		
		This indicates that quantum statistics is conditioned by the topology of the classical multidimensional configuration space of identical indistinguishable particles. Moreover, it has been demonstrated that quantum statistics of the same classical particles can be changed by some external topology-type factors like a strong magnetic field perpendicular to planar system of interacting electrons.
		The origin of such a behaviour is related with the modification of the structure of braid group and next of its 1DURs. Electrons on a planar manifold create at $T=0$ K the triangular (hexagonal) lattice of classical Wigner crystal due to their Coulomb repulsion. As braids are multi-strand trajectories which exchange positions of particles, they could be defined exclusively in the case when braid sizes fit perfectly to positions of electrons in Wigner web. Magnetic field causes in 2D a finite size cyclotron trajectories and the braids are of the similar size as no other trajectories exist at magnetic field presence. Strong enough magnetic field can cause that braids obligatory built of pieces of cyclotron orbits are too short to match even closest electrons in the planar Wigner web. As was proved in \cite{pra} (by application of Bohr-Sommerfeld rule to multiply connected space) multiloop orbits and related generators $\sigma_i^{2k+1}$ ($k$ positive integer) can match neighbouring electrons in the Wigner lattice if ordinary singleloop orbits and related braids $\sigma_i$ were too short. In this way there appear cyclotron subgroups of the genuine full braid group with new generators $\sigma_i^{2k+1}$ instead of $\sigma_i$ and with distinct their 1DURs defining new quantum statistics \cite{laughlin2}. Inclusion of the possibility for matching by multiloop braids also next-nearest neighbouring electrons in the Wigner crystal results in the whole hierarchy of cyclotron subgroups (and related different quantum statistics), which perfectly elucidates experimentally observed fractional quantum Hall effect hierarchy \cite{laughlin2,tsui1982,annals2021}. 
		
		In this example we see that quantum statistics assigned by 1DURs of admitted cyclotron subgroups are apparently flexible to trajectory topology changed by the magnetic field (or Berry field in the case of Chern topological insulators \cite{hasan2010}) and also by electrical field vertically applied to multilayer Hall systems (in bilayer graphene the vertical electrical field can block inter-layer tunnelling of electrons, which precludes hopping of trajectories between layers and eventually changes the statistics of carriers, what is visible in the experiment reported in \cite{maher}). Fractional quantum Hall effect demonstrates that quantum statistics of the same electrons can vary in response to external topological type factors. 
		
		Though in 3D spatial manifolds fractional Hall effect does not exist since cyclotron braids in 3D have an arbitrary size (due to drift motion along magnetic field direction), other topological constraints on braid groups and their unitary representations can be imposed. In the present paper we argue that an extremely strong gravitational field, as induced in the black hole beneath and close beyond the Schwarzschild horizon, can definitely restrict topology of trajectories in 4D spacetime, which completely prohibits the braid group definition and locally washes out quantum statistics. The related proof is presented in Appendix \ref{bbb}.
		
		Note that it is possible to comment on quantum statistics without invoking to Feynman path integrals. 
		Equivalently, the 1DUR of a particular braid defines a phase shift of the multiparticle wave function $\Psi(\mathbf{r}_1,\dots ,\mathbf{r}_N)$ when its arguments $\mathbf{r}_1, \dots, \mathbf{r}_N$ (classical coordinates of particles on the manifold $M$) mutually exchange themselves according to this braid \cite{sud,imbo}. Let us emphasize that the exchanges of coordinates are  not permutations in general, and the paths of exchanges are important, unless the manifold $M$ is a three or higher-dimensional space without linear topological defects, such as strings \cite{birman,sud}. However, the definition of statistics in any dimension unavoidably needs the possibility to implement a braid group, which is conditioned by the existence of classical trajectories for the exchanging of particle positions on the manifold $M$. If such trajectories are prohibited the quantum statistics cannot be defined. This is the case in the Schwarzschild zone beneath the innermost unstable circular orbit  of a  black hole as is shown in Appendix \ref{bbb} for the isotropic not rotating and not charged Schwarzschild black hole.

		\section{Homotopy of geodesics  in the vicinity of the event horizon in Schwarzschild metric}
		
		\label{bbb}
		
		The  Schwarzschild metric for a classical black hole (non-rotating and non-charged) assumed as point-like singularity with the mass $M$  has the form for  line element $ds=c d\tau$ for proper time $\tau$, i.e., as in Eq. (\ref{metryka1})	\cite{schwarzschild},
		\begin{equation}
			\label{schwarzschildmetric}
			-ds^2=-(1-2u)dt^2+(1-2u)^{-1}dr^2+r^2(d\theta^2+sin^2\theta d\phi^2), 
		\end{equation}
		where $u=\frac{GM}{r}$  ($c$ is here assumed 1, $G$ is the gravitation constant, $r,\theta, \phi$ are the spherical coordinates).
		Let us consider a path (in chosen $\theta=\frac{\pi}{2}$ geodesic plane)  of a test particle with unit mass in the Schwarzschild metric.  There are  following constants of the motion,
		\begin{equation}
			k=(1-2u)t^{(1)},\;\;\; l =r^2\phi^{(1)},
		\end{equation}
		where superscript $(1)$ denotes differentiation with respect to proper time. $k$ and $l$ correspond to the kinetical energy and angular momentum of the test particle, respectively. For the radial distance, the following condition can be written,
		\begin{equation}
			(r^{(1)})^2=k^2-1+2u-(1-2u)l^2u^2.
		\end{equation}
		From the above it follows that the smallest stable circular orbit is at $u=\frac{1}{6}$ for $l=2\sqrt{3}$ and $k=\sqrt{\frac{8}{9}}$. The circular orbit for photon is unstable at $u=\frac{1}{3}$ (photon sphere). 
		
	A particle placed beneath the Schwarzschild surface ($u=\frac{1}{2}$) has a dynamics completely controlled by the central singularity and this particle tends to the singularity along a spiral towards the origin.  The same holds for photons and massive particles below the photon sphere ($u=\frac{1}{3}$, i.e.,  $r=\frac{3}{2}r_s$, $r_s=\frac{2GM}{c^2}$) if they pass this sphere inward  -- they  also unavoidably spiral onto Schwarzschild horizon (for a remote observer they spiral onto the event horizon infinitely long although within a finite proper time they pass the event horizon and fall onto central singularity, which is visible in e.g., Novikov \cite{novikov} or  Kruskal–Szekeres coordinates \cite{kruskal,szekeres}). No other trajectories exist for these particles near the Schwarzschild horizon and beneath the photon sphere which is simultaneously the sphere with the radius of innermost unstable circular orbit for massive particles. 
		
		The qualitative change of the trajectory homotopy at the innermost unstable circular orbit is visible in Fig. \ref{photosphere} and can be rationalized as follows.
		The planar (as usual in spherically symmetric gravitation field) trajectory  in the geodesic plane $\theta=\frac{\pi}{2}$ for a particle  with the mass $m$ can be determined by the solution of Hamilton-Jacobi equation,
		\begin{equation}
			g^{ik}\frac{\partial S}{\partial x^i}\frac{\partial S}{\partial x^i}-m^2c^2=0,
			\end{equation}
		which for the Schwarzschild metric  (\ref{metryka1}) attains the form,
		\begin{equation}
			\label{jacobi1}
			\left(1-\frac{r_s}{r}\right)^{-1}\left(  \frac{\partial S}{c\partial t}\right)^2-\left(1-\frac{r_s}{r}\right)\left(\frac{\partial S}{\partial r}\right)^2
			-\frac{1}{r^2}\left(\frac{\partial S}{\partial \phi}\right)^2-m^2c^2=0.
			\end{equation}
		Function $S$ has the form,
		\begin{equation}
			\label{jacobi}
			S=-{\cal{E}}_0t+{\cal{L}}\phi+S_r(r),
			\end{equation}
		where the energy ${\cal{E}}_0$ and the angular momentum ${\cal{L}}$ are constants of motion. By the substitution of Eq. (\ref{jacobi}) to Eq. (\ref{jacobi1}) one can find $\frac{\partial S_r}{\partial r}$ and after integration, 
		\begin{equation}
			S_r=\int dr\left[ \frac{{\cal{E}}_0^2}{c^2}\left(1-\frac{r_s}{r}\right)^{-2}-\left(m^2c^2
			+\frac{{\cal{L}}^2}{r^2}\right)\left(1-\frac{r_s}{r}\right)^{-1}\right]^{1/2}.
			\end{equation}
		The function $r=r(t)$ of the particle trajectory is given by the condition $\frac{\partial S}{\partial{\cal{E}}_0}=const.$,
		which leads to the formula, 
		\begin{equation}
			\label{promien}
			ct=\frac{{\cal{E}}_0}{mc^2}\int \frac{dr}{(1-\frac{r_s}{r})\left[\left(\frac{{\cal{E}}_0}{mc^2}\right)^2-\left(1+\frac{{\cal{L}}^2}{m^2c^2r^2}\right)\left(1-\frac{r_s}{r}\right)\right]^{1/2}}.
			\end{equation}
		The angle-dependence of the trajectory is given by the condition $\frac{\partial S}{\partial{\cal{L}}}=const.$, resulting in the equation,
		\begin{equation}
			\phi=\int dr\frac{{\cal{L}}}{r^2}\left[\frac{{\cal{E}}_0^2}{c^2}-\left(m^2c^2+\frac{{\cal{L}}^2}{r^2}\right)
			\left(1-\frac{r_s}{r}\right)\right]^{-1/2}.
			\end{equation}
		
		Eq. (\ref{promien}) can be rewritten in a differential form,
		\begin{equation}
			\frac{1}{1-r_s/r}\frac{dr}{cdt}=\frac{1}{{\cal{E}}_0}\left[{\cal{E}}_0^2-U^2(r)\right]^{1/2},
				\end{equation}
			with the effective potential,
			\begin{equation}
				U(r)=mc^2\left[\left(1-\frac{r_s}{r}\right)\left(1+\frac{{\cal{L}}^2}{m^2c^2r^2}\right)\right]^{1/2},
				\end{equation}
			where ${\cal{E}}_0$ and ${\cal{L}}$
are energy and angular momentum of the particle, respectively. The condition ${\cal{E}}_0\geq  U(r)	$ defines an accessible region for the motion. ${\cal{E}}_0=  U(r)	$ defines circular orbits.
 Limiting circular orbits are  defined by the  extrema of $U(r)$: maxima define unstable orbits, whereas minima stable ones (depending on parameters ${\cal{E}}_0$ and ${\cal{L}}$, i.e., on integrals of the motion). The conditions $U(r)={\cal{E}}_0$ and $U'(r)=0$ (' denotes the derivative with respect to $r$) have the form,
\begin{equation}
	\label{branches}
	\begin{array}{l}
	{\cal{E}}_0={\cal{L}}c\sqrt\frac{2}{rr_s}\left(1-\frac{r_s}{r}\right),\\	\frac{r}{r_s}=\frac{{\cal{L}}^2}{m^2c^2r_s^2}\left[1\pm \sqrt{1-\frac{3m^2c^2r_s^2}{{\cal{L}}^2}}\right],\\
		\end{array}
	\end{equation}
where the sign $+$ corresponds to stable orbits (minima of $U(r)$) and the sign $-$ to unstable ones (maxima of $U(r)$). From the above it follows that the innermost stable circular orbit is at $r=3r_s$, ${\cal{L}}=\sqrt{3}mcr_s$ and ${\cal{E}}_0=\sqrt{\frac{8}{9}}mc^2$ (point $P$ in Fig. \ref{photosphere} which presents both branches of  the solution of the second equation (\ref{branches})). The innermost unstable circular orbit is at $r=1.5 r_s$ for ${\cal{L}} \rightarrow \infty $ and ${\cal{E}}_0\rightarrow \infty$ (dotted horizontal asymptotic in colour blue in Fig. \ref{photosphere}). This is visualized in Fig. \ref{photosphere} -- the upper curve (red one) gives positions of stable circular orbits and terminates in  the point $P$ at $r=3r_s$ and the lower curve (blue one) gives positions of unstable circular orbits. Below $1.5 r_s$ (being also the radius of the photon sphere, i.e., the sphere with the radius of the smallest unstable orbit for massless particles) none circular orbits for any particle exist and geodesics unavoidably spiral one-way towards the horizon for infalling particles no matter how high the initial energies and angular momenta of  particles are. 

This is a qualitative change of trajectories at $r=1.5 r_s$. The disappearance of circular trajectories (assuring in principle particle exchanges) and the obligatory one-way falling down onto the horizon of  particles passing inward the sphere $r=1.5r_s$, do not allow to construct braids for these particles. Braids must be closed loops in the multiparticle configuration space $F_N=(M^N-\Delta)/S_N$ and for the local manifold $M$ -- the curved space between  the sphere of innermost unstable  circular orbit and any other sphere closer to the horizon, no such loops exist. This precludes exchange positions of these particles in this region because of the absence of suitable trajectories. 

The  stationary Schwarzschild coordinates are convenient to describe in the ordinary rigid coordinates (as in remote system) $t,r,\theta,\phi$ the outer region with respect to the event horizon, thus the region between the innermost unstable circular orbit and the horizon. In these coordinates any reformulation of path integral (\ref{path}) is not required.  The similar trajectory property as below the innermost unstable circular orbit holds, however, also for the inner of a black hole beneath the horizon, where  absolutely all particles one-way spiral to the central singularity, and also no other trajectories exist there. This movement can be conveniently parametrized in nonstationary Kruskal-Szekeres or Novikov metrics, though the homotopy class of trajectories is the  same in all curvilinear coordinates. In curvilinear  coordinates the path integral (\ref{path}) must be reformulated e.g., to proper time and Novikov radial coordinate \cite{novikov} or similar non-stationary and non-rigid coordinates in Kruskal-Szekeres metric \cite{kruskal,szekeres}, which, however, does not change the reasoning related to homotopy of trajectories expressed by the braid group.

	\begin{figure}
		\centering
		\includegraphics[width=0.95\columnwidth]{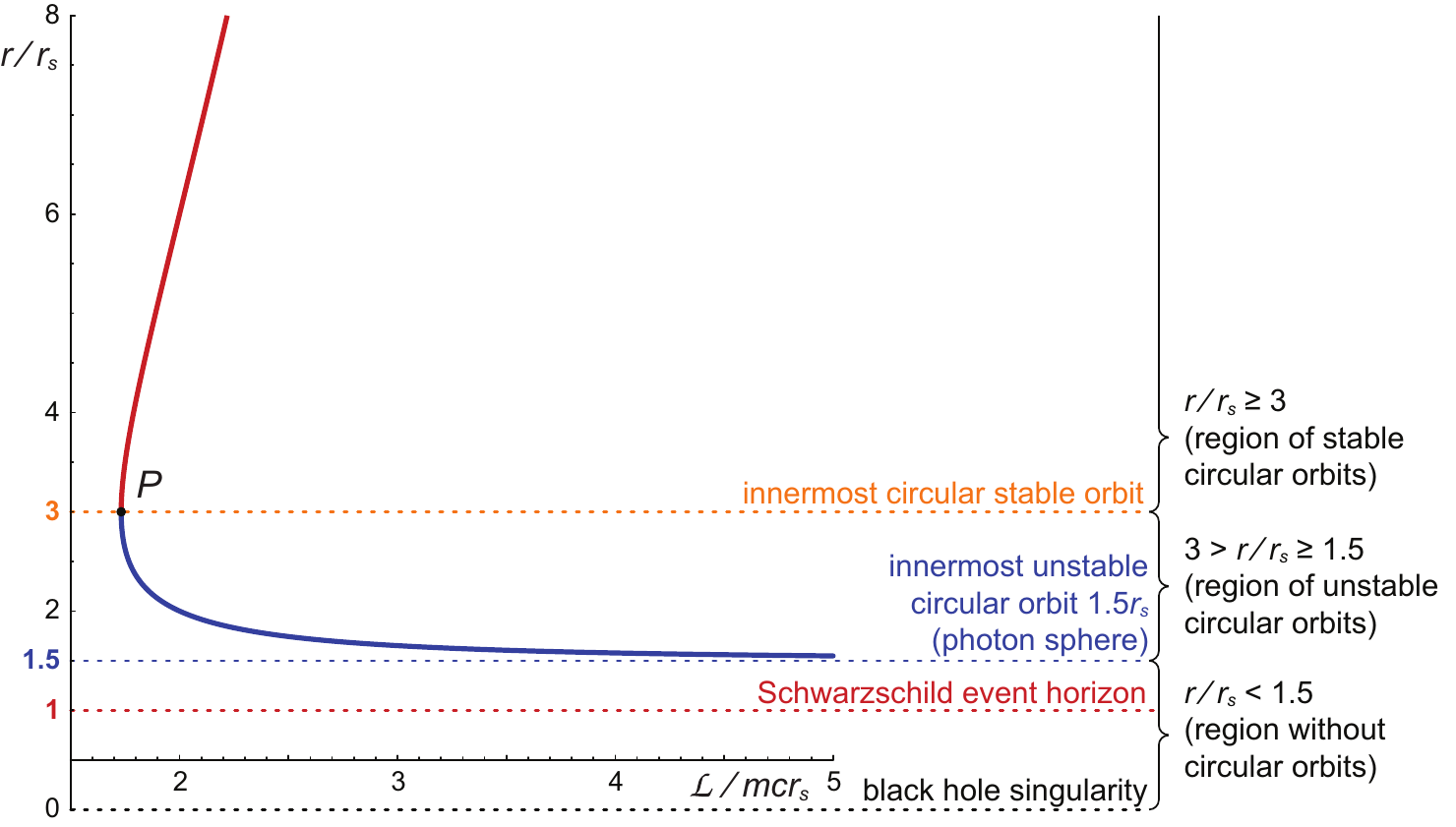}
		\caption{\label{photosphere} Radii of stable (red) and unstable (blue) circular orbits in Schwarzschild geometry. The innermost circular stable orbit occurs at $r=3r_s$ (point $P$ on the level of dotted line indicated in orange colour), whereas the innermost unstable circular orbit occurs at $r=1.5 r_s$ (asymptotic dotted line indicated in blue colour). Below $r=1.5r_s$ none circular orbits exist.  	}
	\end{figure}

		 The unavoidable falling down of trajectories (along spirals in general) of all particles despite their mutual interaction, precludes the definition of the braid group in the local region beneath the innermost unstable circular orbit, because loops in $F_N$ linking distinctly numbered particles do not exist here. The braid group is the collection of nonhomotopic classes of closed loops assuming indistinguishability of particles. These loops must be taken from the domain of all trajectories in the multiparticle configuration space $F_N=(M^N-\Delta)/S_N$. Because of the division by the permutation group $S_N$ this space is not intuitive and can be only imagined as related to distributions of particles equivalent with respect to particle renumbering (differently numbered particle configurations are in fact the same point in $F_N$, which is unintuitive). To cope with such a limitation of an intuitive visualization of braids, the real trajectories in $M$ linking various particles at some fixed but arbitrary  their numeration must be available. If they are not available, the braid group does not exist. The existence of circular orbits assures trajectory topology sufficient to particle interchange -- such trajectories may serve for exchanging of particles located on the same circular orbit and thus, due to indistinguishability, for all particles. Two particles located at diagonal of a circle (i.e., on ends of an arbitrary  diameter of a circle) exchange mutually their positions along such circular trajectory (as illustrated in Fig. \ref{gc}). The existence of circular trajectories assures in topological sense the possibility of the braid group implementation.  The closure of circles of single particle orbits does not mean the closure of loops in $F_N$ but some pieces of circular orbits with constant radius allow for the organization of elementary braids (the generators of the braid group \cite{birman,mermin1979aaa}), which are $N$-thread trajectory bunches exchanging only the $i$-th particle with the $(i+1)$-th one when the other particles remain at rest, at some fixed but arbitrary particle numbering.    In the case of only on-way directed spiral trajectories   of  particles passing inward  the innermost unstable circular orbit, the braid loops linking various particles infalling onto the horizon or next onto central singularity cannot be organized, as the homotopy of geodesics (conserved in all coordinate systems describing the same spacetime curvature of the black hole) precludes the existence of looped trajectories in $F_N$ with $M$ confined to the region from the event horizon up to the innermost unstable circular orbit. The same holds for the inner of the event horizon. Though in various curvilinear coordinate systems (in different metrics for the same gravitationally curved spacetime) the trajectories can be deformed and variously parametrized, their topological homotopy class is conserved. If in one coordinate system (e.g., in Schwarzschild metric) do not exist trajectories for particle exchanges in some region, also in other coordinate system (metric) this property holds, locally precluding the braid group definition, despite various parametrization of geodesics. The local absence of the braid group  causes the local disappearance of quantum statistics and local washing-out of Pauli exclusion principle for half spin particles beneath the innermost unstable circular orbit.

		\begin{figure}
			\centering
			\includegraphics[width=0.95\columnwidth]{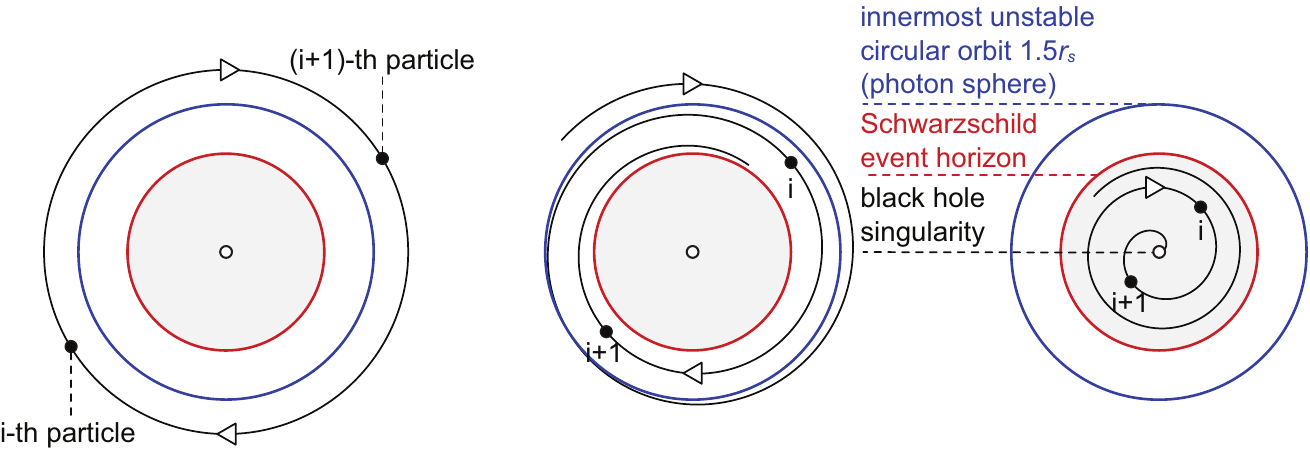}
			\caption{\label{gc} Simplified pictorial illustration of the change of trajectory homotopy at passing the innermost unstable circular orbit of a black hole. When circular orbits are available then the particle position interchange is possible, in principle (left). When only spiral one-way trajectories are admitted and particles unavoidable fall towards the event horizon (in Schwarzschild coordinates (centre))   or central singularity (right) particles cannot mutually interchange  positions. Even if the $i$-th particle  can substitute the position of the other $(i+1)$-th one, the inverse trajectory does not exist no matter how close particles are.	}
		\end{figure}

		 In Schwarzschild geometry the trajectory analysis in the region between the event horizon and the innermost unstable circular orbit can be done in ordinary remote rigid coordinates $(t,r,\theta,\phi)$, however, the inner spatial-volume  of the horizon is zero in this metric. The inner space beneath the event horizon it is better to represent in Kruskal-Szekeres or Novikov metrics, because the rigid static  coordinate system simultaneously for the outer and inner of the horizon does not exist, what is, however, not the restriction  for time-dependent metrics, like the Kruskal-Szekeres metric \cite{kruskal,szekeres} or Novikov metric \cite{novikov}. In these metrics particles smoothly pass the event horizon and spiral into singularity within a finite proper time period (e.g., the proper time $\tau$ is the explicit new time-coordinate in the Novikov metric, whereas the radial coordinate changes to new time-dependent position variable $R$ according to the $\tau$-dependent transformation, $\frac{\tau}{2M}=(R^2+1) \left[\frac{r}{2M}-\frac{(r/2M)^2}{R^2+1}\right]^{1/2}+(R^2+1)^{3/2}arccos\sqrt{\frac{r/2M}{R^2+1}}$, (at $c,G=1$) and $\theta$ and $\phi$ coordinates remain unchanged). The change to other curvilinear coordinates does not disturb, however, the homotopy class of trajectories.





\providecommand{\href}[2]{#2}\begingroup\raggedright\endgroup



\end{document}